# A Study on Software Metrics and its Impact on Software Quality

[1]J.Rashid, [2]T.Mahmood, [3]M.W.Nisar

[1,2]*Computer Science Department, University of Engineering and Technology Taxila, Pakistan*
[3]*Computer Science Department, COMSATS Wah Cantt.*

[1]junaidrashid062@gmail.com

**Abstract-** Software metrics offer a quantitative basis for predicting the software development process. In this way, software quality can be improved very easily. Software quality should be achieved to satisfy the customer with decreasing the software cost and improve there liability of the software product. In this research, we have discussed how the software metrics affect the quality of the software and which stages of its development software metrics have applied. We discussed the different software metrics and how these metrics have an impact on software quality and reliability. These techniques have been used for improving the quality of software and increase the revenue.

**Keywords-**Data stream mining, In-process metrics, defect severity, Test coverage, McCabe, STREW-H, Software Quality, In-process Measurements, Testing Metrics, software complexity.

## I. Introduction

Two things discussed here are the software metrics and software quality. Software metrics are the measuring techniques for systems software, its different parameters or different processes of that software which provide measurements of a different aspect of the software. The measurements are obtained when these metrics techniques are applied so these metrics are the measurements. Software quality is the degree to which the software meets specified requirements of the customer, at the agreed cost within the agreed timescale with the efficient working and effective deliverables. To compete with today's global community, improvement of software development process and software product is needed [1].

The purpose of this survey is to evaluate software metrics and their role in improving software quality, by studying and analyzing the history, methodologies and the future trends of software metrics.

In software development, quality is the main issue which customers always require. Customer expectation for software quality is increasing with the software market. So, the countries and the companies are investing more and more money, effort and time for the improvement of the software quality [2].

Without software development process knowledge, it is quite difficult or nearly impossible to achieve software quality. For good quality software defects, errors or bugs should be identified at the initial stage of the development process else it will become very expensive to handle them if found late [3-5].

Software metrics provide the quantitative measure of the attributes exhibited by the structure of the software and the mathematical measure which is sensitive to the software characteristics differences [6]. Software metrics always useful for managing and controlling the software development process. Its enhance the software quality i.e. they should be simple, clear, easily understandable, definable, reasonable, robust, valid and have some objectives. It is often difficult to capture the complexity of the software which is only possible by the proper use of software metrics because they can capture its complexity which is embedded in program structure of that software [7]. Software metrics are divided into three types: process metrics, product metrics, and project metrics.

Software measurements are also helpful in software configuration management or version because they let us know about which part of the software is modified or changed [7]. That is why software metrics are beneficial in accessing the software's quality [8].

The objectives of this conductive study are given as below.

- How and what metrics should be used to measure software complexity?
- What metrics methodologies are being used by different standards?
- Which metrics are more helpful in achieving software quality and which are less helpful?
- Which does metrics predict software defects at an early stage so that its quality may not affect?
- How to manage quality in complex software?
- What are the impacts of different quality metrics on software and its quality?

Section 2 explain the literature review. In section 3 software metrics and its impact on software quality are discussed. Section 4compares different software metrics on its quality. Section 5 is discussion and section 6 conclude the paper.

## II. LITERATURE REVIEW

History of software metrics starts from the late1960s. In the start LOC, i.e. lines of code, the measure was used for measuring both program quality and the programmer's productivity. The early resource prediction models also used lines of code or related models, e.g. "deliverable source instruction" as key size variable. In1961, software quality metrics were published for the first time, i.e. a number of defects per KLOC. It was used for measuring program complexity" [9]. For the first time, LOC was used in 1974 and then metrics for measuring software complexity was proposed by McCabe in 1976 [10].

In 2000, there was a significant increase in the level of the use of software metrics which was not there before the 1990s. In 1976, which was the start of the use of metrics, complexity metrics were widely used for quality control purpose. In 1994, some attempts were made to make software effective (i.e. Quality software) to count the software defects, but those were not considered as "metrics program" but were simply regarded as "software engineering practices". In 1996, despite their poor size, measure LOC metrics were routinely used as a measure of software quality. Before the 21$^{st}$ century, some researches were made to explain that how a prediction of defects was so much important in quality prediction, in which it was concluded that size, measure or complexity metrics cannot provide an accurate prediction of the software defects alone. They also concluded that traditional statistical methods (i.e. Regression-based methods) were inappropriate for defect prediction.

The requirements of the customer should be properly understood because this is the initial step and its chances to affect the quality are greater than any other phase, but the complexity of this phase is quite less. Various studies show that 25-40% defects are due to the errors in requirement phase. In another place, Ray Ruby study shows that incomplete requirement specifications result in 28% of defects of the software [9]. The above studies show that the color and the requirements specification can improve software quality. Table 1 shows that with the increase of complexity software quality is decreases.

TABLE I  QUALITY VS COMPLEXITY

|  | Complexity | Quality |
|---|---|---|
| Phase 1 | 0.5 | 4.3 |
| Phase 2 | 1 | 4 |
| Phase 3 | 2 | 3.5 |
| Phase 4 | 2 | 3 |
| Phase 5 | 2.5 | 2.5 |
| Phase 6 | 3.5 | 3.5 |

To understand and minimize the complexity of the software many organizations adopt some techniques or methodologies and software metrics is one of them which is being used by almost all organizations. These are the measurement techniques used to check if the software is functional, reliable, usable, efficient, portable, and maintainable which tells about the extent of its quality. They are used for assessment of the quality of the software during and after its development. Their usage will provide quantitative measures for making good decisions about the software quality [11]. Metrics are also used to detect code redundancy, which can be removed by applies refactoring techniques [8].

## III. SOFTWARE METRICS AND ITS IMPACT ON SOFTWARE QUALITY

Software metrics are further classified into three categories as shown in Fig. 1.

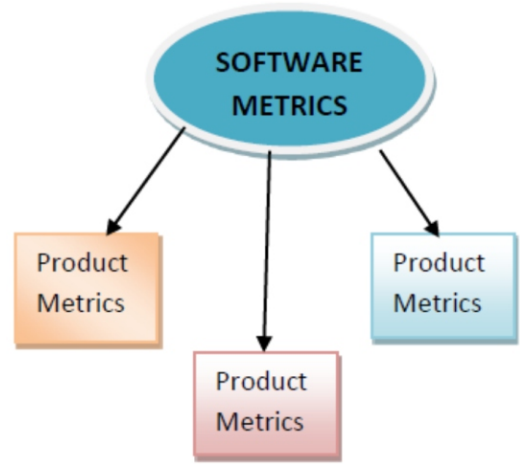

Fig.1. Software metrics

- Product metrics – These are the characteristics of a product, such as its performance, design, size, quality level, and complexity.
- Process metrics – These metrics are used to make better software development and maintenance processes.
- Project metrics – These metrics describe the project execution and its characteristics.

As our center is to see the effect of programming measurements on quality, we will center around programming quality measurements. Programming quality measurements are additionally sorted into item quality, in-process quality, and upkeep, quality as venture quality measurements (estimate, cost, calendar, abilities, and authoritative structure) is less related than the item and process quality.

Along with product and process, maintenance is the main factor affecting the software's quality.

In 2002, according to Pearson, these three-quality metrics are further subdivided into four metrics each as shown below in Table 2.

TABLE II

TYPES OF SOFTWARE QUALITY METRICS

| Software | Quality | Metrics |
|---|---|---|
| **Product quality metrics** | **In-process quality metrics** | **Maintenance quality metrics** |
| Mean time to failure | Defect Density During Machine Testing | Fixture backlog and backlog management index |
| Defect density | Defect Arrival Pattern During Machine Testing | Fixture backlog and backlog management index |
| Customer problems | Phase-Based Defect Removal Pattern | Percent delinquent fixes |
| Customer satisfaction. | Defect Removal Effectiveness | Fix quality |

*A. Product quality metrics*

What's powerful product exceptional metrics that are getting used to enhance the product great?

Records movement mining strategies are used to are expecting software program effects on the premise of its source code metrics [12]. Building predictive models are of excellent benefit. Constructing it using records flow mining strategies is better because it doesn't have big everlasting garage requirements for simplest stores as a quick and limited quantity of build exchange units that is due to large garage requirements [13-14].

The consulted research has offered a solution for encoding software program satisfactory metrics as records streams. In the case of Jazz after execution of software program build, the facts streams could be supplied, the student copies this sort of facts movement that's taken from ancient information [15]. The model which generated from software program build histories, we will run the real-time facts [13][16].

The consequences have shown that data stream mining strategies preserve lots capacity as the Jazz surroundings. Predictive models may be up to date and encoded into the IDE as a tool while construct is executed. real-time remarks can be provided to the contributors through this device through the progress of code in relative to the metrics expected and extracted build outcome. For a team to communicate effectively with the successful generation of bold, this tool will provide real-time insight for good communication. Data stream mining has applications in ATM transactions and safety, web searches, traffic systems, managing the network and networks/data [17-18].

*B. In-process metrics*

For an early indication of an external measure of defect density, what is easy to measure, internal in-processing methods?

Based on Nagappan's study [19], they propose STREW-H method which uses a suite of internal, in-process metrics which estimates defect density metric suite. Some metrics are not as applicable to functional languages, because of different language paradigms e.g. LOC [20] is a most commonly used metric which easily gathers with either paradigm, where class structure metrics are not relevant to size measurements. In his study, STREW-J was used to start.

The best proposed in-process metric method was STREW-H. The proposed metric suite is refined by deleting and adding metrics until it is felt that a minimal set of metrics needed to accurately explain and predict product defect density has achieved [21].

These methods are helpful for software engineers to detect the defect very soon and remove these defects.

*C. Maintenance quality metrics*

Several software metrics techniques are used to achieve quality through its maintenance and performance. Functional metrics measure the amount of business functionality by measuring the size (Through LOC and FP) and budget [22].Periodicity metrics are used for higher planning, scheduling, and periodic protection sports. where making plans develops procedures and scheduling evaluates the supply of the sources required for maintenance of the software project in the exact time [23]. In 2010, Meselhy introduced two impartial periodicity assets because of preservation, the pre-planned protection and failure repair Preventive [24].OEE metrics or equipment performance metrics: In 1988, Nakajima launched the concept of the total productive maintenance (TPM) which provides a quantitative metric called "Overall Equipment Effectiveness (OEE)" which measures the productivity of manufacturing equipment [25]. It is the most popular equipment performance indicator which is measured by different production losses [26]. In 2006. Marquez defined maintenance process as the series of action at different stages [27]. The maintenance process has two parts efficiency analysis and effective analysis. The first part identifies the suitable procedure and the second part helps to detect the important problems and locate their potential solution. The researchers have defined eight phases to accomplish this assessment as:

1. Factory performance.
2. Software quality.
3. Effective study for solutions.
4. Maintenance performance analysis.
5. The action plan.
6. Action plain implementing.
7. Action monitor.
8. Adoption of plans [24].

## TABLE III  COMPARISON OF METRICS

| Metrics Type | Problem | Technique | Solution | Result |
|---|---|---|---|---|
| **Product quality metrics** | What are effective PQM which is being used to improve the product quality? | • Data stream mining techniques<br>• Building predictive models | Encoding software quality metrics as data streams are used. | Data stream mining techniques hold much potential as Jazz environment |
| **In-process quality metrics** | What is IPM for measuring defect density? | • The STREW-J technique was used to start.<br>• STREW-H technique. | Best proposed in-process metric method for predicting defect density is STREW-H. | The strew-h method provides early warnings which are helpful for the early prediction of defect density. |
| **Maintenance quality metrics** | What metrics techniques are used to achieve quality through its maintenance and performance and how? | • Functional metrics.<br>• Periodicity metrics.<br>• Maintenance process metrics.<br>• Equipment performance metrics<br>• (OEE) metrics. | The literature review of (MPIs) maintenance performance indicators is divided into five categories as shown in the techniques. | There is a strong dependency between maintenance performance metrics (MPMs) and the maintenance performance indicators (MPIs). |

There is a strong dependency between maintenance performance metrics (MPMs) and the maintenance performance indicators (MPIs). Table 3 shows the product quality metrics, in process quality and maintenance quality metrics.

The recent studies show that companies derived metrics as four further categories:

- Quality Metrics
- Productivity and Schedule Metrics
- Assessment Metrics
- Business and Corporate Metrics

From these four categories, we will see what techniques and metrics are further used in quality metrics, as our purpose is to focus on achieving quality using metrics. Quality metrics are further divided into the metrics shown below in figure 2:

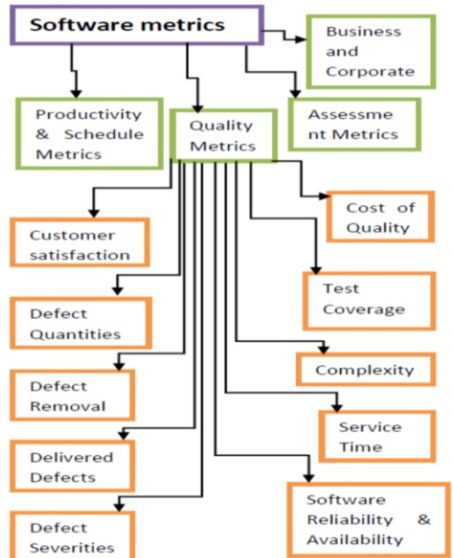

Fig.2. Types of software metrics

### D. Customer satisfaction

Which model defines customer satisfaction factor and what are the suitable measurable variables in this model?

Customer satisfaction is an important factor for the success of different companies and is the way to achieve quality products [28]. Seven hypothetical variables define (ECSI) European customer satisfaction index model as shown in figure 3.

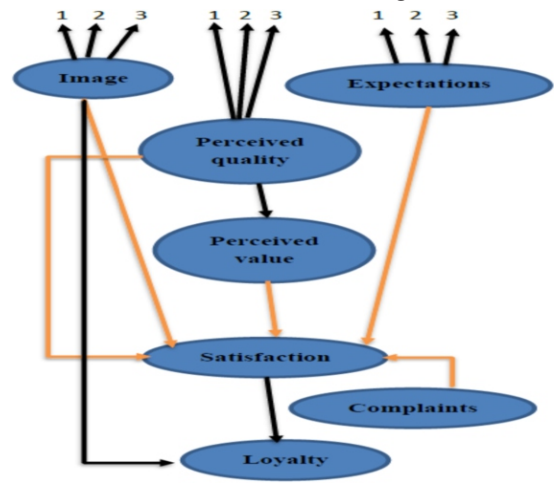

Fig.3. Model ECSI (European customer satisfaction index

The different hypothetical variables are discussed in [29] [30].

1. Image
2. Customer Expectation
3. Perceived Quality of Service/Product.
4. Perceived Value
5. Customer Satisfaction
6. Customer Complaints
7. Loyalty

These variables are used in the calculation of the Customer Satisfaction Index (CSI) as shown in equation 1 below.

$$\epsilon_j = \frac{\sum_{i=1}^{n} v_{ij} . x_{ij}}{10 \sum_{i=1}^{n} v_{ij}} \tag{1}$$

The above equation1 measures the results of customer satisfaction and the indexes of the variables and is based on how the customer evaluates the product and its services, therefore, proved to be the important metric for achieving software quality by measuring the value of customer satisfaction which is an important factor for the quality product [30]. The results of the calculations can be used by the client for decision making related to the purchase of goods and services. It is also used by the firm as it creates the firm development strategy [29].

### E.  Defect Quantities

What are the possible software defect quantities and how to detect those defects?

Defect quantities are the bugs, errors or the defects observed in all artifacts of the system development life cycle, i.e. requirements, design, code, testing documents and secondary defects or "bad fixes" [31]. Table 4 shows the average number of defects found in software projects [32].

TABLE IV
DEFECT ORIGIN AND DEFECT POTENTIALS

| Defect Origin | Defect Potentials |
|---|---|
| Requirements | 1.00 |
| Design | 1.25 |
| Coding | 0.75 |
| Documents | 0.60 |
| Bad fixes/Secondary Defects | 0.40 |
| Total | 5.00 |

Figure 4 shows the increase and the decrease of the defects found during different phases of the system development life cycle, i.e. requirements, design, coding, documents and "bad fixes".

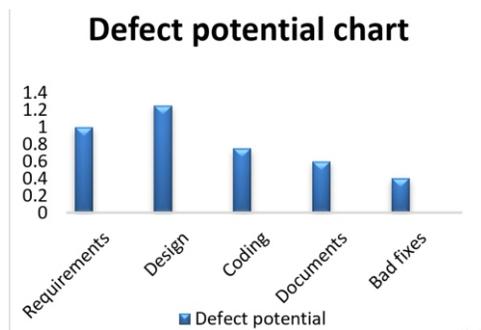

Fig.4. Defect potential chart

Defects per function point range below 2.00. They vary with CMM levels for other factors [31].

The bad fixes percentage is calculated as given in [33]. Equation 2 describes the bad fixes percentage.

$$Bad\ defect\ \% = \frac{Total\ valid\ defects}{Bad\ fix\ defects} * 100 \tag{2}$$

### F.  Defect removal

How to reap excessive defect elimination efficiency?

The elimination metrics illness defects are determined and eliminated earlier than the product is brought to the consumer. So, to obtain an excessive disorder elimination performance formal inspection and formal trying out are finished [31]. table five shows the proportion of defects removed before the product delivery to the clients [32].

TABLE V
DEFECT ORIGIN AND REMOVAL

| Defect Origin | Defect removal efficiency |
|---|---|
| Requirements | 77% |
| Design | 85% |
| Coding | 95% |
| Documents | 80% |
| Bad fixes/Secondary Defects | 70% |
| Total | 85% |

### G.  Delivered defects

Defects are the variety of supply in keeping with FP or LOC at the time of delivery of the product requirements disorder, design disorder, code illness, documentation/online assist defect, defect added by means of fixes, and so on. In result of the above study, table 6 shows the number of each of these defects and its average found.

TABLE VI
DEFECT ORIGIN AND DELIVERED

| Defect Origin | Delivered defects |
|---|---|
| Requirements | 0.23 |
| Design | 0.09 |
| Coding | 0.19 |
| Documents | 0.12 |
| Bad fixes/Secondary Defects | 0.12 |
| Total | 0.75 |

Figure 5 shows the increase and the decrease of the delivered defects during different phases of the system development life cycle, i.e. requirements, design, coding, documents and "bad fixes".

### H.  Defects Severities

Defects severity level is the degree of end user's business impact which affects the quality of the software. This severity level of the defects can be

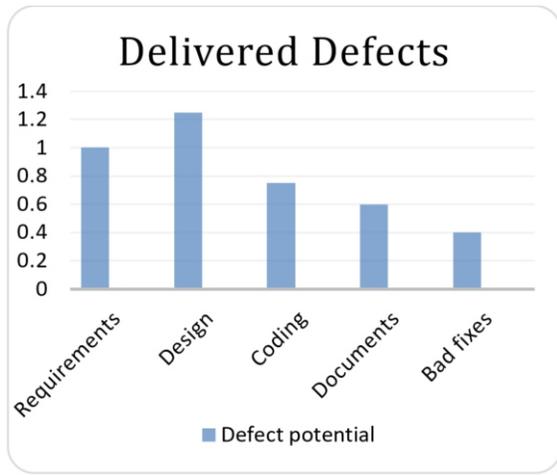

Fig.5. Delivered defects chart

determined by software testing. Business impact = effect on the end user x frequency of occurrence

High defect severity => low product quality

By determining defect severity, we can minimize it to enhance product quality. Every defect contains its severity level which can be low, medium or high. It is important to measure the severity of the product because it helps

- In determining the efficiency of the test process.
- In deciding the priority of defects
- In making the bug tracking process effective
- In making release decisions

Classification of defect severity:

- Fetal defects
- Major defects
- Minor defects
- Cosmetic defects

Defect severity can be measured using the defect severity index, a metric used to measure the quality of the product directly. Four levels are maintained to check defects severity i.e.

Level 4: Critical, level 3: Serious, level 2: Medium and level 1: Low

From [33], a study shows the defect severity index as shown in equation 3.

$$Bad\ severity = \frac{severity * valid\ defects}{total\ defects} \quad (3)$$

Table 4, table 5 and table 6 shows that if the average defect potential is 5.00 errors or defects per FP i.e. function point and the average percentage of defect removal efficiency is 85%, then the total number of the delivered defects will be around 0.75 defects per FP i.e. function point. The summary of these tables giving some concluding remarks.

1) The largest number of defects are more likely to occur (Table 4)
2) Defect removal efficiency is not very good i.e., observed 85% where it should be about 95%. (Table 5)
3) Delivered defects originate of different types from multiple sources (Table 6)

Figure 6 shows a comparison of defect potentials and delivered defects.

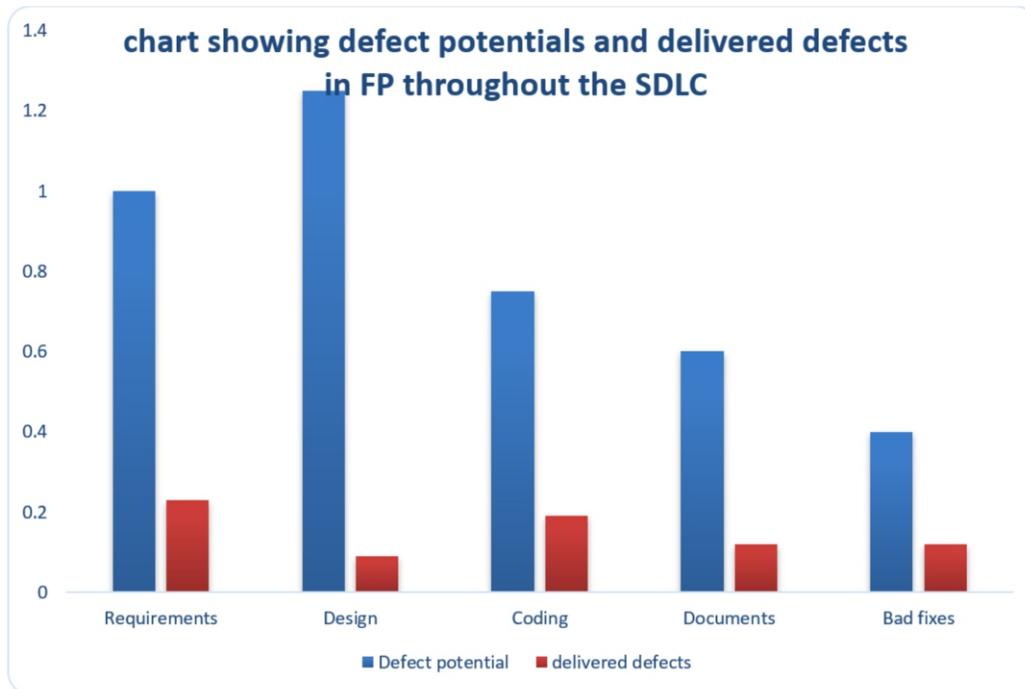

Fig.6. Chart showing defect potentials and delivered defects in FP throughout the SDLC

*Software Reliability & Availability*

What is software reliability metrics and how they are helpful in measuring reliability when software sizes have no uniform definition and it is very difficult to understand the nature of the software?

Software reliability and availability are the main factors which cannot be ignored because they are responsible for gaining the customer's satisfaction and is also difficult to measure [34].

Reliability is the measure of the probability of the software that for how long it will work and will not fail. The unreliability of the software is caused due to the failure or the bugs or the design faults. The measure of software reliability the reliability metrics are used. They express the reliability of the software product quantitatively. Which metric is to be used is decided on the basis of the type of the system to which it is applied. The quality of software depends on the factors including software reliability model and software quality metrics [35].

Figure7 shows the software quality improvement factors.

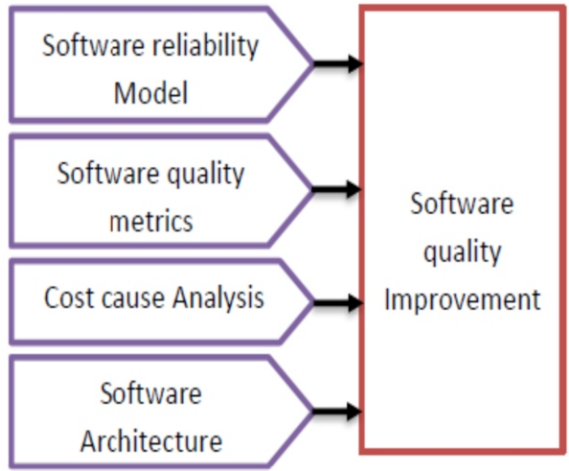

Fig.7.Software quality improvement factors

Without having a good understanding of the software's nature, it is quite difficult to measure the software's reliability. Therefore, instead of measuring the reliability directly, some reliability metrics are used which reflects reliability related characteristics of the software product [35]. These are discussed below.

• Mean time to failure (MTTF): it's miles the time interval among successive failures [36].

• Meantime among failures (MTBF): it is the aggregate of the 2 metrics MTTF and MTTR to offer MTBF.

• The price of the prevalence of a failure (ROCOF): The call suggests it's for the number of screw-ups going on in line with unit time which is also known as failure depth metric.

• Mean time to repair (MTTR): This metric degree the time taken to monitoring the mistake and then

solving it. Availability (AVAIL): As described above, it is the probability of the software or the system to be present or available when it is needed [35].

Different software reliability metrics in different phases of SDLC are

• Requirement reliability metrics

• Design & code reliability metrics

• Testing reliability metrics [37].

Figure8 shows the software reliability metrics.

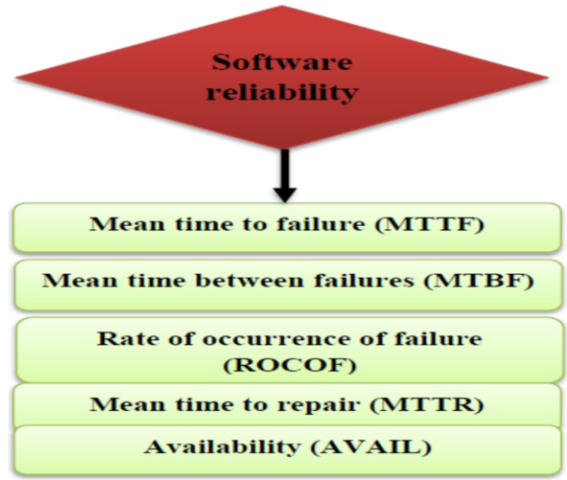

Fig.8. Software reliability metrics

The achieving of software reliability is the key task of any company. It doesn't only tell about the current reliability of the software but also forecasts about its future reliability. Thus, using various measurement techniques of the software mentioned here in the paper, to eliminate any error or fault of the software process, that is how it improves the reliability of the software product [34-35].

**Service Time/ Response time/ defect turnaround time**

Turnaround time or the response time for defect fixes, by the level of severity [38]. To measure the turnaround time following equation 4 is used.

$$Defect\ turn\ time = \frac{Actual\ time\ fix\ defects}{Plain\ time\ fix\ defects}\ (4)$$

**Software complexity:**

Software complexity is an essential factor which should always be taken into consideration along with software parameters. Software cost increases with the increase of the complexity and the decrease of the reliability [39].

The research in [40-41] shows that software complexity metric is further categories as shown in figure 9 below.

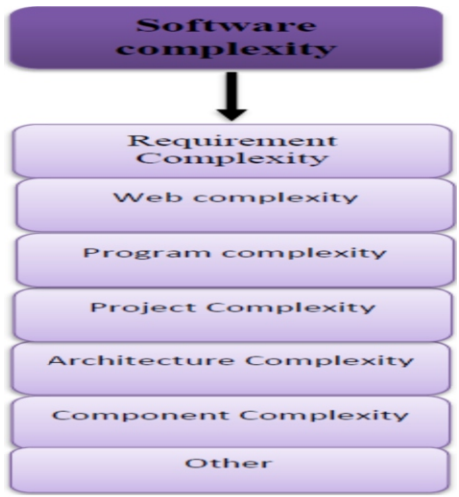

Fig.9. Software complexity framework

There are some metrics to calculate the complexity of the software. The study of different complexity metrics, the metrics which are used in the development life cycle are shown below in Table 7.

TABLE VII
COMPLEXITY METRICS AND THEIR PHASES

| Reliability Metrics | Phases |
|---|---|
| McCabe | Design, coding |
| Halstead | Deployment, |
| Line of code Error count | Coding, deployment, Maintenance |
| OOP class metrics | Design, coding, |
| Software package metrics | Design, deployment, |
| Cohesion | Design |
| Coupling | Design |

Figure 10 shows the Software complexity categories.

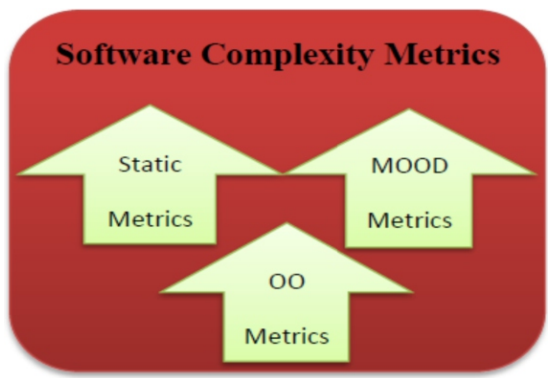

Fig. 10. Software complexity categories

Table 8 shows the three-complexity metrics and their sub-categories along with their complexity levels [41-46].

TABLE VIII
METRICS COMPLEXITY

| Category | Metric Name | Complexity |
|---|---|---|
| **Static metrics** | Source Line of Code | High-Complexity |
| | Comment Percentage [42] [43] [44] | Low-Complexity |
| | Halstead Metrics | High-Complexity |
| | Maccabees Cyclomatic Complexity | High-Complexity |
| | Weighted Method Per Class [45] | High-Complexity |
| **Object-Oriented-Metrics** | Depth of Inheritance Tree | High-Complexity |
| | Number Of Children | High-Complexity |
| | Coupling Between Object Class | High-Complexity |
| | Response Of A Class | High-Complexity |
| | Lack Of Cohesion In Methods | Low-Complexity |
| | Method Hiding Factor | High-Complexity |
| | Attribute Hiding Factor | High-Complexity |
| **Metrics of object-oriented design (MOOD)** | Method Inheritance Factor | Low-Complexity |
| | Attribute Inheritance Factor | Low-Complexity |
| | Polymorphism Factor | Low-Complexity |
| | Coupling Factor | Low-Complexity |

For an item, oriented complexity metrics as C&ok technique and mood are getting used for the closing two decades [47]. it is impossible to make present-day systems without object-orientated design and object-oriented programming. The OO design carries all properties for all small and large initiatives which complements software exceptional [48]. In [49] shows that they build a unified software program complexity metric which relies on different dimensions, domains, and factors of software program size. The following goals are useful for software.

1.   It gives a dimension which permits contrast of the unique relative complexities of two unique packages
2.   It returns the space among the two arguments via taking values from an issue and measuring the gap.
    It estimates the productiveness of the people in software tasks to obtain indirect measures [49].
3.   A set of quality KPIs was selected carefully with broad acceptance in the development team to improve the quality of complex software projects [50].

Test Coverage
    It is the measure of the number of tests performed by the set of the tests. Use a fully automated Test Coverage Analysis Tool for measuring test coverage [51].

Cost of Quality:

*Principles of Quality, Costs*, illustrates a technique for analyzing the quality, cost by breaking down these costs [52]. The cost of quality shown in equation 5.

$$\text{Quality cost} = \text{Conformance cost} + \text{nonconformance} \quad (5)$$

Where,

Cost of conformance = Appraisal cost + prevention cost

And,

Cost of Nonconformance = Cost of internal failure + Cost of external failure

From the above formula, a study was performed to calculate the cost of quality as shown in table 9.

TABLE IX
MEASURE OF THE TOTAL COST OF QUALITY

| Cost of quality | Non-formal testing | Automated testing | Manual testing |
|---|---|---|---|
| Conformance cost | | | |
| Nonconformance cost | $0 | $70,000 | $82,500 |
| The total cost of Quality | $752,500 | $437,500 | $302,500 |
| | $752,500 | $507,500 | $385,000 |

By the recent study performed by Anuradha K, she applied different test metrics to check the quality of the software.

Table 10 gives an overview of the methodologies discussed in this paper for above-discussed software quality metrics.

TABLE X
METHODOLOGIES FOR SOFTWARE QUALITY METRIC

| QUALITY METRICS | METHODOLOGIES DISCUSSED IN PAPER |
|---|---|
| Customer satisfaction | (ECSI) European customer satisfaction index model<br>Customer satisfaction index<br><br>$$\varepsilon_j = \frac{\sum_{i=1}^{n} v_{ij} \cdot x_{ij}}{10 \sum_{i=1}^{n} v_{ij}}$$ |
| Defect Quantities | Defect potentials found by the origin of defects circa 2009 in terms of defects per FP |
| Defect removal | Defect removal efficiency found by the origin of defects circa 2009 |
| Delivered defects | Defects delivered found by the origin of defects circa 2009 in terms of defects per FP<br><br>Defect severity index= |
| Defect Severities | Σ (Severity Index x No Of Valid Defects)<br>----------------------------------------<br>Total Number of Valid Defects |
| Software Reliability & Availability | • Mean time to failure (MTTF)<br>• Mean time between failures (MTBF)<br>• The rate of occurrence of a failure (ROCOF)<br>• Mean time to repair (MTTR)<br>• Availability (AVAIL)<br>• Requirement reliability metrics<br>• Design & code reliability metrics<br>• Testing reliability metrics |
| Service Time/Response time | Defect Turnaround time =<br>(Actual time was taken to fix the defect) / (Planned time is taken to fix the defect) |
| Software complexity | Static complexity metrics<br>Object-oriented complexity metrics (OO metric) and<br>Metrics for object-oriented design (MOOD) |
| Test Coverage | Test Coverage=<br>(Number of coverage items exercised/ Total number of coverage items) x 100 |
| Cost of Quality | Cost of quality=<br>Cost of conformance + Cost of nonconformance |

The purpose of this paper is to observe the impact of different software metrics on the quality of software i.e. software quality metrics impact. Therefore, here we compare these metrics in table 11 below, to see whether their impact is low or high and what are the reasons for these impacts.

## IV. Comparison of Software Metrics on its Quality

Software quality is an essential thing for any organization and its depend that how the customers are satisfied with software product [53]. Software quality is maintained with some standards and procedures of a software organization. Several organization follows different standards for software quality. The software quality product also depends on the effective requirement engineering process [54] and other applications [55].

## V. Discussion

The metrics described in the paper are used for providing accurate reports on the daily or weekly basis. These metrics proved to be very useful for the assessment of project status and its quality. ECSI model should be followed to fulfill customer satisfaction. Reliability of the product should always be taken in the concern, for that purpose, reliability metric should be

TABLE XI
COMPARISON OF METRIC IMPACT ON SOFTWARE QUALITY

| SR # | REF # | PROBLEM | SOLUTION | TECHNIQUE | RESULT | IMPACT |
|---|---|---|---|---|---|---|
| 01 | [xii-xviii] | What are effective PQM which is being used to improve the product quality? | Encoding software quality metrics as data streams are used. | • Data stream mining techniques<br>• Building predictive models | Data stream mining techniques are better to use in product quality metrics because it holds much potential as the Jazz environment. | Data mining has a high impact compared to the jazz environment. |
| 02 | [xix-xxi] | What is IPM for measuring defect density? | The proposed in process metric method for predicting defect density in STREW-H | • The STREW - J technique was used to start<br>• STREW-H technique. | The strew-h method provides early warnings which are helpful for the early prediction of defect density. | The strew-h method has a high impact on software quality as compare to STREW-J technique. |
| 03 | [xxv] | What metrics techniques are used to achieve quality through its maintenance and performance and how? | The literature review of (MPIs) maintenance performance indicators is divided into five categories as shown in the techniques. | The techniques are given below:<br>• Functional metrics.<br>• Periodicity metrics.<br>• Maintenance process metrics.<br>• Equipment performance metrics<br>• (OEE) metrics. | High dependency on the performance of metrics. | The techniques for MPI, all prove to be good to measure maintenance for quality. So have a high impact on software quality. |
| 04 | [xxviii-xxx] | Which model defines customer satisfaction factor and what are the suitable measurable variables in this model? | ESCI model defines Image Customer, Expectation Perceived, Quality of Service/ Product, Perceived Value, Customer Satisfaction, Customer Complaints Loyalty | (ECSI) European customer satisfaction index model defines seven hypothetical variables as given in the solution. Generally, the customer satisfaction index is used as shown in formula 8 below. Formula 8: $$\varepsilon_j = \frac{\sum_{i=1}^{n} v_{ij} \cdot x_{ij}}{10 \sum_{i=1}^{n} v_{ij}}$$ | Customer satisfaction is an important factor for the success of different companies and is the way to achieve quality products | ECSI model helps in determining product quality through determining customer's satisfaction. So, give a high impact on software quality. |
| 05 | [xxxi-xxxiii] | What are the possible software defect quantities and how to detect those defects? | Defect quantities are the bugs errors or the defects observed in all artifacts of the system development life cycle i.e. requirements, design, code, testing documents and secondary defects or "bad fixes". These measured in terms of function points. | Defect potentials found by Origin of Defects Circa 2009 in terms of defects per FP as shown in formula 9. Formula 9: Bad Fix Defect%= Total Number of Valid Defects -------*100[%] Number of Bad Fix Defects | The largest number of defects are more likely to occur | Defect quantities have a high impact on software quality as it evolves through the development life cycle and predicts the occurrence of a number of defects for different phases. |
| 06 | [xxxi-xxxii] | How to achieve a high defect removal efficiency and what is its impact on software quality? | To achieve a high defect removal efficiency formal inspection and formal testing are done. | Defect Removal Efficiency found by Origin of Defects Circa 2009 in terms of defects per FP | Defects removal efficiency is not very good i.e., observed 85% where it should be about 95%. | High impact as high defect removal efficiency leads to a quality product. |
| 07 | [xxxi-xxxii] | How to measure defect density of delivered defects and what is its impact on software quality? | Defect density metric gives the defect density measure. | Defect density metrics Defect density= Defects/ unit size Where unit size is in term of a number of lines of code. | Delivered defects originate of different types from multiple sources. They keep on writing. | Low impact on software quality as Large no of delivering defects gives low product quality |

| SR # | REF # | PROBLEM | SOLUTION | TECHNIQUE | RESULT | IMPACT |
|---|---|---|---|---|---|---|
| 08 | [xxxiii] | How to measure defect severity level and what is its impact on software quality? | This severity level of the defects can be determined by software testing. The Defect Severity Index is used for calculating defect severity. | The formula 10 shows that. Formula 10: Defect Severity Index = $\Sigma$ (Severity Index x No of Valid Defects) ---------------- Total Number of Valid Defects | High defect severity => low product quality | Low impact on product quality as High defect severity => low product quality. |
| 09 | [xxxiv-xxxvii] | What are the impact of defect software reliability and availability measure on the quality of software? How is it calculated? | Metrics shown in techniques are used to measure software reliability and availability. | Mean time to failure (MTTF) Mean time between failures (MTBF) The rate of occurrence of failure (ROCOF) Mean time to repair (MTTR) Availability (AVAIL) | Using these reliability techniques we can eliminate any error or fault of the software process that is how it improves the reliability of the software product. | High impact as it helps in improving software so improves product quality. |
| 10 | [xxxviii] | What is the impact of defect turnaround time on the quality of software? How is it calculated? | It is measured by calculating the actual and planned time for fixing defects. | The formula 11 shows. Formula 11: Defect Turnaround time = (Actual time was taken to fix the defect)/ (Planned time is taken to fix the defect) | Can only be calculated when the fix is successfully delivered. | Has a high impact on product quality. |
| 11 | [xxxix--xlv] | What is the impact of software complexity on its quality? Which complexity metrics have high impact? | The set of quality KPIs was selected carefully with broad acceptance in the development team to improve the quality of complex software projects. | Static complexity metrics Object-oriented complexity metrics (OO metric) and Metrics for object-oriented design (MOOD) Which includes many like Maccabees Cyclomatic Complexity = E-N+P and Halstead Metrics | Complexity increase => reliability decreases. | Quality decrease as complexity increases, so it has a low impact on software quality. |
| 12 | [11] | How to measure test coverage? What is its impact on software quality? | Use a fully automated Test Coverage Analysis Tool | Test Coverage= (Number of coverage items exercised a / Total number of coverage items) x 100 | Coverage can lead away from the real goal of testing. | It has a low impact on product quality. |
| 13 | [12] | How to measure the cost of quality? What is its impact on software quality? | We can calculate it through the cost of conformance and nonconformance. | Cost of quality= Cost of conformance + Cost of nonconformance Where Cost of conformance = Appraisal cost + prevention cost and Cost of nonconformance= Cost of internal failure + Cost of external failure | Predicts the budget. | Has a high impact on product quality. |

used in different phases of development of the software product, i.e. McCabe, Halstead, a line of code, error count, Object-oriented programming class metrics, software package metrics, cohesion and coupling metrics. MTTF and MTTR metrics are should also be used to measure the availability and reliability of the software product. Object-oriented complexity metrics prove to be very good in predicting complexity fast and, therefore, these metrics enhance software quality.

## VI. CONCLUSION

The paper presents the results of an initial methodical attempt to calculate build success or failure of the software product by using source code metrics.

Data stream mining technique is better for measuring product quality.

The automated tools are currently being created to gather the information and provide it to developers automatically while they are still coding and can inexpensively make actions to correct the mistakes. The initial viability analysis was performed by using a subset of metrics from the STREW-H for this purpose.

The complexity metrics are the best among all three categories of complexity metrics.

It is best to use the quality metrics obtained from the advanced study to improve the software's quality.

To increase the quality of the software product reliability metrics, complexity metrics, defect removal efficiency, etc. help in gaining quality & customer satisfaction model index is best for gaining customer satisfaction which leads to the quality of the software product.

## REFERENCES

[1] L. Lazic, N.Mastorakis,"Cost-effective software test metrics", WSEAS Transactions on Computers. 2008 Jun 1;7(6):599-619.

[2] B. Kitchenham ,SL. Pfleeger SL, "Software quality: the elusive target [special issues section]", IEEE Software. 1996 Jan;13(1):12-21.

[3] N. Nagappan ,L. Williams , M. Vouk ,J. Osborne ,"Early estimation of software quality using in-process testing metrics: a controlled case study", InACM SIGSOFT Software Engineering Notes 2005 May 17 (Vol. 30, No. 4, pp. 1-7). ACM.

[4] BW. Boehm,"Software engineering economics", Englewood Cliffs (NJ): Prentice-hall; 1981 Oct 22.

[5] "J.E. Gaffney, JR, Metrics in software quality assurance", ACM '81 Proceedings of the ACM '81 conference, 1 January 1981, Pages 126-130.

[6] W. Li,H. Delugach ,"Software metrics and application domain complexity", InSoftware Engineering Conference, 1997. Asia Pacific... and International Computer Science Conference 1997. APSEC'97 and ICSC'97. Proceedings 1997 Dec 2 (pp. 513-514). IEEE.

[7] K. Stroggylos, D. Spinellis , "Refactoring--Does It Improve Software Quality? ", InSoftware Quality, 2007. WoSQ'07: ICSE Workshops 2007. Fifth International Workshop on 2007 May 20 (pp. 10-10). IEEE.

[8] T. Mens, S. Demeyer. "Future trends in software evolution metrics". InProceedings of the 4th international workshop on Principles of software evolution 2001 Sep 10 (pp. 83-86). ACM.

[9] NE. Fenton,M. Neil, "Software metrics: successes, failures and new directions", Journal of Systems and Software. 1999 Jul 1;47(2-3):149-57.

[10] T. Honglei ,S. Wei ,Z. Yanan , " The research on software metrics and software complexity metrics", InComputer Science-Technology and Applications, 2009. IFCSTA'09. International Forum on 2009 Dec 25 (Vol. 1, pp. 131-136). IEEE.

[11] Z. Markov, I. Russell, "An introduction to the WEKA data mining system", ACM SIGCSE Bulletin 38 (3) (2006) 367–368.

[12] Z. Markov, I. Russell, "An introduction to the WEKA data mining system", ACM SIGCSE Bulletin 38 (3) (2006) 367–368.

[13] B. Pfahringer, G. Holmes, R. Kirkby, "Handling numeric attributes in hoeffding trees", in T. Washio et al. (Eds.), Advances in Knowledge Discovery and Data Mining, Springer, Berlin Heidelberg, 2008, pp. 296–307.

[14] A. Bifet, "Adaptive learning and mining for data streams and frequent patterns", ACM SIGKDD Explorations Newsletter 11 (1) (2009) 55–56.

[15] J. Finlay, A.M. Connor, R. Pears, "Mining Software Metrics from Jazz", in Software Engineering Research, Management, and Applications, 2011 9th International Conference on, 2011.

[16] K.J. Cios et al., "Data Mining: A Knowledge Discovery Approach", Springer, Publishing Company, Incorporated, 2010.

[17] J. Finlay, R. Pears, A. M. Connor:" Data stream mining for predicting software build outcomes using source code metrics", Journal of Information and software technology, 2013.

[18] G. Hulten, L. Spencer, P. Domingos, "Mining time-changing data streams", in Proceedings of the Seventh ACM SIGKDD International Conference on Knowledge Discovery and Data Mining, ACM, San Francisco, California, 2001, pp. 97–106

[19] N. Nagappan , L. Williams , M. A Vouk,, "Towards a metric suite for early software reliability assessment" InInternational Symposium on Software Reliability Engineering, FastAbstract, Denver, CO 2003 Nov (pp. 238-239).

[20] TM. Khoshgoftaar, JC. Munson, "The lines of code metric as a predictor of program faults: A critical analysis", InComputer Software and Applications Conference, 1990. COMPSAC 90.

Proceedings., Fourteenth Annual International 1990 Oct (pp. 408-413). IEEE.

[21] M. Sherriff,L. Williams,M. Vouk , "Using In-Process Metrics to Predict Defect Density in Haskell Programs", InFast Abstract, International Symposium on Software Reliability Engineering, St. Malo, France 2004 Nov 2.

[22] R. Meli:, "Functional Metrics: Problems and Possible Solutions.", 2013.

[23] J. Kumar, V. Soni,G. Agnihotri, "Maintenance performance metrics for manufacturing industry", International Journal of Research in Engineering and Technology. 2013;2(2):136-42.

[24] K. T. Meselhy , W,.H. ElMaraghy ,"A periodicity metric for assessing maintenance strategies", CIRP Journal of Manufacturing Science and Technology. 2010 Jan 1;3(2):135-41.

[25] S. Nakajima, "Introduction to TPM: Total Productive Maintenance (preventative maintenance series)", Hardcover. ISBN 0-91529-923-2. 1988.

[26] P. Muchiri , L. Pintelon ,"Performance measurement using overall equipment effectiveness (OEE): literature review and practical application discussion", International journal of production research. 2008 Jul 1;46(13):3517-35.

[27] A. C. Marquez,J. N. Gupta, "Contemporary maintenance management: process, framework and supporting pillars", Omega. 2006 Jun 1;34(3):313-26.

[28] V. Chalupský, "Marketingový audit spokojenosti zákazníků.", VUTIUM; 2001.

[29] K. Ryglová, K., and I. Vajčnerová. "Potential for utilization of the European customer satisfaction index in Agra-business." 2005 (4): 161–167.

[30] F. Dařena F, A. Motyčka , R. Malo, "Customer satisfaction index calculation service", InConference 2008 Sep (Vol. 151, pp. 24-26).

[31] C. Jones, " Measuring defect potentials and defect removal efficiency" , The Journal of Defense Software Engineering. 2008 Jun;21(6):11-3.

[32] C. Jones "Software Quality and Software Economics", October 19, 2009.

[33] L. LAZIC, "Software Quality & Testing Metrics." management 4: 6.

[34] G. K. Saha , " Software Reliability Issues: Concept Map", IEEE Reliability Society 2009 Annual Technology Report.

[35] G. Kaur G, K. Bahl, "Software reliability, metrics, reliability improvement using agile process", IJISET-International Journal of Innovative Science, Engineering & Technology. 2014 May;1(3):143-7.

[36] P. Ramachandran P, SV. Adve , P. Bose, J.A Rivers, J. Srinivasan ,"Metrics for lifetime reliability", 2006.

[37] V. Tiwari,R. K. Pandey, "Open source software

and reliability metrics", International Journal of Advanced Research in Computer and Communication Engineering. 2012 Dec;1(10):808-15.

[38] K. Sakthi, and R. Baskaran. "Defect analysis and prevention for software process quality improvement." International Journal of Computer Applications 8.7 (2010).

[39] S. Yu , S. Zhou S, " A survey on metric of software complexity", InInformation Management and Engineering (ICIME), 2010 The 2nd IEEE International Conference on 2010 Apr 16 (pp. 352-356). IEEE.

[40] C. F. Kemmerer,"Software complexity and software maintenance: A survey of empirical research", Annals of Software Engineering. 1995 Dec 1;1(1):1-22.

[41] R. Jiang, "Research and Measurement of Software Complexity Based on Wu Li, Shili, Renli (WSR) and Information Entropy." Entropy 17.4 (2015): 2094-2116.

[42] M. Lorenz, and J. Kidd, "Object-oriented software metrics: a practical guide",  Prentice-Hall, Inc., 1994.

[43] R. C. Sharble and S. C. Samuel , "The object-oriented brewery: a comparison of two object-oriented development methods." ACM SIGSOFT Software Engineering Notes 18.2 (1993): 60-73.

[44] K. A. Ferreira,M. A. Bigonha , R. S. Bigonha , L. F. Mendes , H.C. Almeida ,"Identifying thresholds for object-oriented software metrics", Journal of Systems and Software. 2012 Feb 1;85(2):244-57.

[45] A. H. Watson, and D. R.  Wallace. "Structured testing: A testing methodology using the cyclomatic complexity metric, "NIST (National Institute of Standards and Technology), special Publication 500.235 (1996): 1-114.

[46] C.  Chawla, G. Kaur "Comparative Study of the Software Metrics for the complexity and Maintainability of Software Development.", (IJACSA) International Journal of Advanced Computer Science and Applications, Vol. 4, No. 9, 2013.

[47] T. Honglei , S. Wei, and Z. Yanan. "The research on software metrics and software complexity metrics." Computer Science-Technology and Applications, 2009. IFCSTA'09. International Forum on. Vol. 1. IEEE, 2009.

[48] D. Arora , P. Khanna , A. Tripathi , S. Sharma , "Software quality estimation through object oriented design metrics", Int. J. Computer Science and Network Security. 2011 Apr;11(4):100-4.

[49] R. R. Gonzalez, "A Unified Metric of Software Complexity: Measuring Productivity, Quality, and Value", 1995, 29:17-37.

[50] K. Schüler , R. Trogus , M. Feilkas , T. Kinnen , "Managing Product Quality in Complex Software Development Projects", InProceedings of the

Embedded World Conference 2015.

[51] H. Zhu,P.A Hall , M. H. May , "Software unit test coverage and adequacy". Acm computing surveys (csur). 1997 Dec 1;29(4):366-427.

[52] R. Black, "Investing in Software Testing: The Cost of Software Quality." Rex Black Consulting (2000).

[53] J. Rashid, M. W. Nisar, 2016. How to Improve a Software Quality Assurance in Software Development-A Survey. International Journal of Computer Science and Information Security,

14(8), p.99.

[54] M. Khan, J. Rashid, M.W. Nisar, 2016. A CMMI Complaint Requirement Development Life Cycle. International Journal of Computer Science and Information Security, 14(9), p.1000.

[55] J. Rashid, W. Mehmood, and M. W. Nisar, "A Survey of Model Comparison Strategies and Techniques in Model Driven Engineering," International Journal of Software Engineering and Technology (IJSET), vol. 1, pp. 165-176, 2016.